\documentstyle[12pt,epsfig]{article}
\def\ovl{\overline}
\def\bra{\langle}
\def\ket{\rangle}
\def\dsps{\displaystyle}
\begin{document}

\vspace{.4cm}
\begin{flushright}
\large{HUTP-97/A038\\
July 1997\\
version 0.1}
\end{flushright}
\vspace{0.5cm}
\begin{center}
\Large{\bf Inclusive $CP$ Asymmetries in 
      Semileptonic Decays of $B$ Mesons\footnote{A talk given
      at the San Miniato conference, April 1997, and at the
      Hawaii CP violation conference, February 1997.}}
\end{center}

\vspace{1.2cm}
\begin{center}\large
Hitoshi Yamamoto\\
\vspace{.4cm}
{\normalsize 
{\it Dept. of Physics, Harvard University, 42 Oxford St., Cambridge, 
MA 02138, U.S.A.}\\
{\it e-mail: yamamoto@huhepl.harvard.edu}}
\vspace{.4cm}
\end{center}

\thispagestyle{empty}
\vspace{3.4cm}
\centerline{\Large \bf Abstract}
\vspace{0.3cm}
We estimate the sensitivity of single lepton $CP$ violation 
measurements with respect to that of traditional di-lepton 
measurements. We find that the sensitivity of the single lepton method 
is better than that of the di-lepton method.
The achievable sensitivity with the currently available 
data is already in the range relevant to standard model predictions.
We also give general expressions for inclusive decay time 
distributions on $\Upsilon 4S$ where
the other $B$ is not measured, which will be used to obtain time
dependent asymmetries. The expression is of general use whenever
one deals with inclusive time-dependent as well as time-integrated
measurements in $\Upsilon(4S)\to B^0\bar B^0$ where the final state of
the other $B$ is not reconstructed or when only the time difference is
measured.

\section{Introduction}

The particle-antiparticle imbalance in the mass eigenstates of
the neutral $B$ meson system,
\[
    \delta\equiv     
      {|\bra B^0|B_{a,b}\ket|^2 - |\bra\ovl B^0|B_{a,b}\ket|^2\over
       |\bra B^0|B_{a,b}\ket|^2 + |\bra\ovl B^0|B_{a,b}\ket|^2}\; ,
\]
can be extracted from the same-sign di-lepton asymmetry in 
$\Upsilon(4S)\to B^0 \ovl B^0$ \cite{Okun+,AliAydin}:
\[
     A_{\ell\ell}\equiv
    { N(\ell^+\ell^+) - N(\ell^-\ell^-) \over
      N(\ell^+\ell^+) + N(\ell^-\ell^-) } 
      \sim 2\delta\; .
\]
The short distance calculation gives \cite{Ma+,Barger+,Hagelin}
\[
     A_{\ell\ell}\sim -\,4\pi{m_c^2\over m_t^2}\,
       \Im\left({V_{cb}V_{cd}^*\over V_{tb}V_{td}^*}\right)
     \sim 10^{-3}\;.
\]
The long distance effects, however, are expected to dominate and
one estimate gives $|A_{\ell\ell}| \sim 10^{-3}$ to $10^{-2}$
\cite{Altomari+}.
Thus, a measurement of such
asymmetry does not determine the $CP$ violating
parameters in the standard model. 
However, an experimental value of 
$|\delta|$ above $\sim10^{-2}$ would signal 
new physics \cite{Asl-Newphys,Newphys}. The current published 
experimental number is not yet in the relevant range:
$A_{\ell\ell} = 0.031\pm0.096\pm0.032$ \cite{CLEO-A}.
For $B_s$, relevant CKM factor is obtained by
replacing $d$ by $s$ in the $B_d$ case, and it is expected to be
small.

The $CP$ asymmetry in single lepton sample 
had been suggested as a possible observable to search for $CP$ violation
in the case when the mixing is small \cite{Azimov,Hagelin,Buras+}.
The logic was that if the mixing is
small, then the statistics of the di-lepton events will decrease,
making the di-lepton method impractical.
After the observation of substantial mixing in the neutral 
$B$ meson system \cite{Bmixobs}, however, the single lepton method 
has not received much attention. 
We point out that the advantage of the single lepton method
over the di-lepton method actually increases for large mixings, and that,
on $\Upsilon(4S)$, the single lepton method has a
better sensitivity than the di-lepton method. This is so in spite of the fact
that in the single lepton measurement, one usually cannot distinguish charged
and neutral $B$ mesons. We first present general expressions the inclusive
decay time distribution on $\Upsilon(4S)$.

\section{Inclusive Time Distributions in $\Upsilon(4S)\to B^0\bar B^0$}

The mass
eigenstates can be written in terms of $B^0$ and $\ovl B^0$ as
\[
   \left\{
   \begin{array}{rcl@{\quad}l}
      B_a &=& p  B^0 + q  \ovl B^0 
            &(\hbox{mass:}m_a,\hbox{decay rate:}\gamma_a)\\
      B_b &=& p' B^0 - q' \ovl B^0 
            &(\hbox{mass:}m_b,\hbox{decay rate:}\gamma_b)\\
   \end{array}, \right.
\]
where 
$|p|^2 + |q|^2 = |p'|^2 + |q'|^2 = 1$.
We will adopt the Wigner-Weisskopf formalism\cite{Wigner+} which is valid when
the oscillations caused by differences of masses and the decay rates
are sufficiently slower than the time scale of decay 
transitions.
If we have a pure $B^0$ or $\ovl B^0$ at $t=0$, 
the decay time distributions to a final state $f$ are given by~\cite{lepasym}
\[
  \begin{array}{rcl}
   \Gamma_{B^0\to f}(t) &=& |c|^2 
        \left[ |q'a_f|^2 e^{-\gamma_a t} +
               |q b_f|^2 e^{-\gamma_b t} +
             2 \Re((q'a_f)^*(q b_f) e^{-(\gamma_+ - i\delta m)t}) \right] ,
           \nonumber \\
   \Gamma_{\ovl B^0\to f}(t) &=& |c|^2 
        \left[ |p'a_f|^2 e^{-\gamma_a t} +
               |p b_f|^2 e^{-\gamma_b t} -
             2 \Re((p'a_f)^*(p b_f) e^{-(\gamma_+ - i\delta m)t}) \right] ,
  \end{array}
\]
where
\[
    c \equiv {1\over p'q + pq'},\;
    \delta m\equiv m_a - m_b\; ,\;
    \gamma_\pm \equiv {\gamma_a\pm\gamma_b\over 2}
\]
and $a_f \equiv Amp(B_a\to f)$, $b_f \equiv Amp(B_b\to f)$
are normalized as $\sum_f |a_f|^2 = \gamma_a$, $\sum_f |b_f|^2 = \gamma_b$.

On $\Upsilon(4S)$, the $B^0 - \bar B^0$ pair is created in a $P$-wave state,
and the quantum correlations of the two decays need to be carefully taken into
account. In particular, when one side is detected as a certain final state $f$ that
both $B^0$ and $\bar B^0$ can decay to, then at the same proper time the
other side becomes a superposition of $B^0$ and $\bar B^0$ which depends on the
decay amplitudes of $B^0$ and $\bar B^0$ to $f$. For the double-time decay
distribution, the prescription is well known and the general expression
for the decay probability where one side decays to $f_1$ at $t_1$ and the 
other side to $f_2$ at $t_2$ is~\cite{lepasym}
\begin{eqnarray}
   \Gamma_{\Upsilon(4S)\to f_1f_2}(t_1,t_2) &=& {|c|^2\over2}
\left[ e^{-\gamma_a t_1 -\gamma_b t_2} |a_{f_1}b_{f_2}|^2 +
       e^{-\gamma_b t_1 -\gamma_a t_2} |b_{f_1}a_{f_2}|^2 
    \right. \label{eq:Upsgen} \\
   &&\left. \; -
   2 \Re\left( e^{-(\gamma_+ - i \delta m) t_1} 
               e^{-(\gamma_+ + i \delta m) t_2} 
      (a_{f_1} b_{f_2})^* (b_{f_1}a_{f_2}) \right)\right],\nonumber
\end{eqnarray}
or in terms of $t_\pm \equiv t_1 \pm t_2$,
\begin{eqnarray}
   \Gamma_{\Upsilon(4S)\to f_1f_2}(t_+,t_-) &=&  \label{eq:Upsgenpm} \\
     &&  \hspace{-3.5cm} {|c|^2\over4}e^{-\gamma_+ t_+}
\left[ e^{-\gamma_- t_-} |a_{f_1}b_{f_2}|^2 +
       e^{ \gamma_- t_-} |b_{f_1}a_{f_2}|^2 -
   2 \Re\left(
      (a_{f_1} b_{f_2})^* (b_{f_1}a_{f_2}) e^{i \delta m t_-}
      \right)\right]. \nonumber
\end{eqnarray}

In order to obtain the inclusive decay distribution where only one decay is
detected, one has to integrate over time and sum over all possible final states
on the other side. Performing the operation in (\ref{eq:Upsgen}), and using the
Bell-Steinberger relation\cite{Bell-Stein}
\[
     {\sum_f a_f^*b_f\over\gamma_+ - i\delta m} =
     \bra B_a | B_b \ket\;\; ( =  p'p^* - q'q^* )\; ,
\]
one obtains
\[
  \begin{array}{l}
   {\dsps \Gamma_{\Upsilon(4S)\to f}(t) =
    2 \sum_{f_1}\int_0^\infty 
      \Gamma_{\Upsilon(4S)\to f_1f}(t_1,t)\, dt_1 = 
      \Gamma_{B^0\to f}(t) + \Gamma_{\ovl B^0\to f}(t) }\\
   = |c|^2 \left[ |a_f|^2 e^{-\gamma_a t} +
                |b_f|^2 e^{-\gamma_b t} -
             2 \Re((p'p^*-q'q^*) a_f b_f^* e^{-(\gamma_+ + i\delta m)t}) \right] ,
  \end{array}
\]
where $\Gamma_{\Upsilon(4S)\to f}(t)$ is the probability density
that one finds a given final state $f$ decaying at time $t$ in the
process $\Upsilon(4S)\to B^0\ovl B^0$, and the factor of two arises from the
fact that the given final state can come from either side. This result may be
expected since there is one $B^0$ and one $\bar B^0$ created and no correlation
is measured, so the inclusive distribution can be expected to be simple sum of
the two distributions. Such argument, however, is not in general true for
general two-body states~\cite{inclusive}. At $B$-factories, the absolute decay time
is not measured well; instead the relevant time is the time difference 
$t_-$ of the two decays. This can be obtained by integrating 
(\ref{eq:Upsgenpm}) over $t_+$ and summing over $f_2$:
\[
 \begin{array}{l}
    \Gamma_{\Upsilon(4S)\to f}(t_-) =
        {\dsps {|c|^2\over 2\gamma_+} }e^{-\gamma_+ |t_-|}\times \\
            \left[
         \gamma_b e^{-\gamma_- t_-} |a_f|^2 + 
         \gamma_a e^{ \gamma_- t_-} |b_f|^2 -
    2 \Re\Big( (\gamma_+ - i \delta m) (p'p^* - q'q^*) 
          a_f b_f^* e^{-i\delta m t_-}
       \Big) \right] .
 \end{array}
\]
The sign of the time difference $t_-$ is defined to be the decay time of 
the final state $f$ minus the decay time of the other side.
These formuli above are valid even when $CPT$ is violated.

\section{Leptonic $CP$ Asymmetries and Experimental Sensitivities}

The decay distributions for semileptonic decays are obtained by
the substitution (assuming $CPT$~\cite{Kostelecky}, 
  and $\Delta B = \Delta Q$~\cite{Sarma})
\[
  \begin{array}{ll}
    a_{\ell^+} = p A_0\;, &  b_{\ell^+} = p A_0\; , \\
    a_{\ell^-} = q A_0\;, &  b_{\ell^-} = -q \bar A_0\; ,
  \end{array}
\]
where $A_0$ is the normalized semileptonic decay amplitude.
The decay time distributions for a single $B^0$ or $\bar B^0$
created at $t=0$ are then
\begin{eqnarray}
  \Gamma_{B^0\to\ell^+}(t) &=& \Gamma_{\ovl B^0\to\ell^-}(t) 
   \nonumber \\ &=&
  {A_0^2\over2} e^{-\gamma_+ t} 
       \left[\cosh\gamma_- t + \cos\delta m t\right]\; , \nonumber \\
  \Gamma_{\ovl B^0\to\ell^+}(t) &=& {|p|^2\over |q|^2}
  {A_0^2\over2} e^{-\gamma_+ t}
       \left[\cosh\gamma_- t - \cos\delta m t\right]\; , \nonumber\\
  \Gamma_{ B^0\to\ell^-}(t)     &=& {|q|^2\over |p|^2}
  {A_0^2\over2} e^{-\gamma_+ t}
       \left[\cosh\gamma_- t - \cos\delta m t\right]\; . \nonumber
\end{eqnarray}
We see that the `right-sign' or `unmixed' decay distributions for
$B^0$ and $\bar B^0$ are identical, but the `wrong-sign' or
`mixed' decay distribution can be different when $|p|\not=|q|$ namely
when there is an imbalance of particle and antiparticle in the
physical states. Note, however, that the shape is the same between
$B^0$ and $\bar B^0$ even
for the unmixed decay distributions.

\begin{figure}
 \centering
 \mbox{\psfig{figure=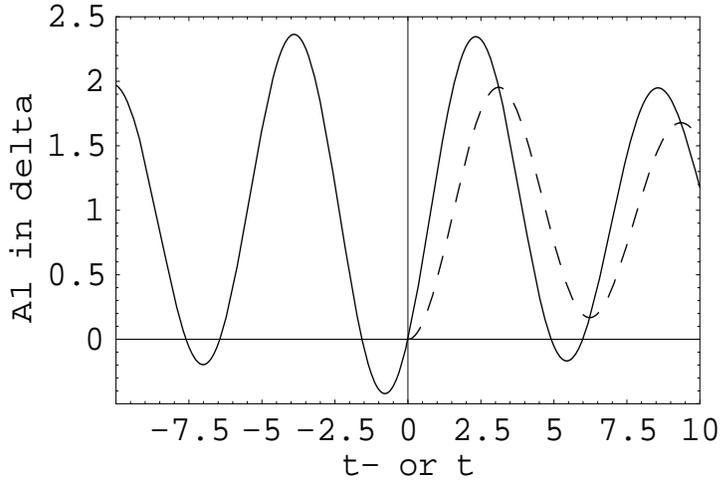,height=2.5in,width=3.7in}}
 \caption{Time-dependent asymmetries of flavor-untagged 
          inclusive semileptonic decay shown in unit of 
          $\delta$. The solid line is the
          asymmetry as a function of decay time difference, and 
          the dashed line is that of absolute decay time.
          Used $\delta m/\gamma_+ = 0.73$ and
          $\gamma_-/\gamma_+ = 0.1$.}
 \label{fig:asymt}
\end{figure}

Flavor-untagged inclusive single lepton decay distribution on
$\Upsilon(4S)$
is then obtained from the
distributions for $B^0$ and $\bar B^0$ by simple incoherent sum. Resulting
time dependent lepton asymmetry is
\[
 \begin{array}{rl}
  A_\ell(t) &= {\dsps {\Gamma_{B^0,\ovl B^0\to\ell^+}(t) -
                  \Gamma_{B^0,\ovl B^0\to\ell^-}(t) \over
                  \Gamma_{B^0,\ovl B^0\to\ell^+}(t) +
                  \Gamma_{B^0,\ovl B^0\to\ell^-}(t)} }\\
            &= {\dsps
    \delta \left(1 - {\cos\delta mt \over\cosh\gamma_- t}\right)\;,}
 \end{array}
\]
which is the asymmetry one would observe in hadronic machines or LEP where
the absolute decay time can be measured. 
The asymmetry starts out as zero at $t=0$ and reaches the
first maximum at around $\delta m t\sim\pi$ (about 4 times the $b$
lifetime). At $B$-factories, the asymmetry
as a function of $t_-$ is
\[
  A_\ell(t_-) =
    \delta \left(1 - {\gamma_+\cos\delta mt_- - \delta m\, \sin \delta mt_-
      \over \gamma_+\cosh\gamma_- t_- + \gamma_- \sinh \gamma_- t_-} \right)\;,
\]
which is a function similar to $A_\ell(t)$ above but the oscillation
amplitude is slightly larger and it applies also to negative value of $t_-$.
Figure~\ref{fig:asymt} shows these asymmetries.

The time-integrated single lepton asymmetry is given by
\[
    A_\ell 
    = 2 D\,\chi\,\delta\;,
\]
where $D$ is the dilution factor due to charged
$B$ mesons, and is equal to the fraction of leptons coming from neutral $B$
mesons. Other dilution effects such as those due to misidentified
leptons or leptons from charmed hadrons could also be absorbed into $D$. 
Assuming that there are the same number of leptons from
charged $B$'s as from neutral $B$'s, we take $D=1/2$.

We will now estimate the sensitivities to $\delta$ of single and
di-lepton asymmetry measurements. We assume that the lepton detection
efficiency $\epsilon_\ell$ for each lepton 
is the same in the single and di-lepton cases,
and that they are uncorrelated in the latter.
Also we assume $\delta\ll 1$ for the expressions of
asymmetries below. In estimating statistics, we further assume
$\gamma_a\sim\gamma_b$ (or equivalently 
$y\ll 1$).
If we have $N_0$ $\Upsilon(4S)\to B^0\ovl B^0$ decays, then
the total number of same
sign di-lepton events detected is $N_0\, b_{sl}^2\,\chi\,\epsilon_\ell^2$,
where $b_{sl}$ is the semileptonic branching fraction. The error
in $\delta$ is then
\[
   \sigma_{\delta}(\ell\ell) = {1\over2}
         {1\over\sqrt{N_0\, b_{sl}^2\,\chi\,\epsilon_\ell^2}}\;.
\]

The total number of single lepton events detected is
$N_0\, 4 b_{sl}\,\epsilon_\ell$; thus the sensitivity to $\delta$
of the single lepton measurement is
\begin{equation}
    \sigma_{\delta}(\ell) = {1\over\chi}
       {1\over\sqrt{N_0\, 4 b_{sl}\,\epsilon_\ell}}\;.
\end{equation}
The ratio of sensitivities of single to di-lepton measurements is then
\begin{equation}
  {\sigma_{\delta}(\ell)\over\sigma_{\delta}(\ell\ell)} =
    \sqrt{{b_{sl}\,\epsilon_\ell\over\chi}}\; .
\end{equation}
We see that the larger the mixing, the more advantageous the single
lepton method becomes. This may be counter-intuitive, but can be
understood as follows: as $\chi$ increases, the
statistics goes up linearly for
the di-lepton sample while its asymmetry stays the same.  
For the single lepton sample, 
the statistics stays the same while the asymmetry goes
up linearly, which is equivalent to statistics increasing quadratically for
a fixed asymmetry.

A typical value for $\epsilon_\ell$ is 0.5. Together with the experimental
values for $b_{sl}$ and $\chi$, the ratio above is 0.78. Namely, the single
lepton measurement has a sensitivity comparable to or better than 
that of the di-lepton
measurement. Note also that the
single and di-lepton datasets are largely statistically independent
(only about 10\%\ of the single lepton events are also in the di-lepton
dataset). The two measurements can thus be combined to improve
overall sensitivity. For example, the current CLEO data corresponds
to $N_0\sim 2\times10^6$. This
gives $\sigma_{\delta}(\ell) = 0.6\%$ and 
$\sigma_{\delta}(\ell\ell) = 0.8\%$ with the combined sensitivity of
0.5\% which is already in the range relevant to 
standard model predictions.
 
When $B^0$'s and $\ovl B^0$'s are generated incoherently (e.g. on
$Z^0$ or in $p\bar p$ collisions), one cannot perform the correlated
di-lepton analysis. However, one
can still perform single lepton asymmetry measurements.
There, alternative method is to use flavor-tagging by lepton and/or jet charge
to enhance the sensitivity.

\section*{Acknowledgments}
This work was supported by the Department of Energy
Grant DE-FG02-91ER40654.

\end{document}